# Aqueye+: a new ultrafast single photon counter for optical high time resolution astrophysics

L. Zampieri*[a], G. Naletto[b,c], C. Barbieri[d], E. Verroi[e], M. Barbieri[f], G. Ceribella[d], M. D'Alessandro[a], G. Farisato[a], A. Di Paola[g], P. Zoccarato[h]

[a]INAF-Astronomical Observatory of Padova, Vicolo dell'Osservatorio 5, 35122 Padova, Italy; [b]Dept. of Information Engineering, University of Padova, Via Gradenigo 6/A, 35131 Padova, Italy; [c]CNR/IFN/LUXOR, Via Trasea 7, 35131 Padova, Italy; [d]Dept. of Physics and Astronomy, University of Padova, Vicolo Osservatorio 3, 35122 Padova, Italy; [e]Centre of Studies and Activities for Space (CISAS) 'G. Colombo', University of Padova, Via Venezia 15, 35131 Padova, Italy; [f]Dept. of Physics, University of Atacama, Copayapu 485, Copiapo, Chile; [g]INAF-Astronomical Observatory of Rome, Via Frascati 33, 00040 Monte Porzio Catone, Rome, Italy; [h]Trimble Terrasat GmbH, Haringstraße 19, 85635 Höhenkirchen-Siegertsbrunn, Munich, Germany

## ABSTRACT

Aqueye+ is a new ultrafast optical single photon counter, based on single photon avalanche photodiodes (SPAD) and a 4-fold split-pupil concept. It is a completely revisited version of its predecessor, Aqueye, successfully mounted at the 182 cm Copernicus telescope in Asiago. Here we will present the new technological features implemented on Aqueye+, namely a state of the art timing system, a dedicated and optimized optical train, a high sensitivity and high frame rate field camera and remote control, which will give Aqueye plus much superior performances with respect to its predecessor, unparalleled by any other existing fast photometer. The instrument will host also an optical vorticity module to achieve high performance astronomical coronography and a real time acquisition of atmospheric seeing unit. The present paper describes the instrument and its first performances.

**Keywords:** Photon counters, High Time Resolution Astrophysics, Optical pulsars

## 1. PHOTON COUNTING DETECTORS FOR OPTICAL HIGH TIME RESOLUTION ASTROPHYSICS

Time is a fundamental variable in Astronomy. However, the time domain below 1 s remains largely unexplored, especially in the visible part of the spectrum which lags behind radio and X-ray wavelengths. A crucial reason is the detector performance. In fact, most optical astronomical instrumentation use charge-coupled devices (CCDs) which are not really suited for optical High Time Resolution Astrophysics (HTRA), since they are usually operated in integration mode. Three instruments specifically designed for investigating sub-ms variability in the optical band have recently been built: ULTRACAM[1] and GASP[2], based on Electron Multiplying CCD (EMCCD) and scientific Complementary Metal-Oxide Semiconductor (sCMOS) technology, respectively, and OPTIMA[3], based on avalanche photodiodes. An optical timing instrument with imaging capabilities is installed, at present, at the SALT telescope and is based on microchannel plates (BVIT[4]) which, however, do not have a high quantum efficiency. Another optical-infrared photon counting instrument, having moderate time resolution (2 microseconds) is ARCONS[5], based on microwave kinetic inductance detectors.

We pioneered the use of Geiger-mode Single Photon Avalanche Photodiodes (SPAD) building Aqueye for the Asiago 1.8 m Copernico telescope[6] and Iqueye for the ESO 3.6 m New Technology Telescope (NTT)[7]. SPADs are, at present, the photon counting detectors with the best timing performance, 30-50 ps time resolution, 10-100 dark counts/s, 6-8 MHz maximum count rate, visible quantum efficiency up to 60%[8,9].

*luca.zampieri@oapd.inaf.it; phone 39 049 829-3433; fax 39 049 875-9840; web.oapd.inaf.it/zampieri

## 2. AQUEYE AND IQUEYE

Aqueye and Iqueye (Figure 1) are the prototypes for a quantum photometer for the future 40 m class telescopes such as the E-ELT[10] and represent the world recognized leading astronomical instruments for the shortest time scales in the optical band[6,7]. They couple the ultra-high time resolution of SPAD detectors with a split-pupil optical concept and a sophisticated timing system. The high quantum efficiency and low temporal jitter, the capability to time tag and store the arrival time of each individual photon with better than 100 ps relative time resolution (< 0.5 ns absolute time accuracy with respect to UTC), and the possibility to bin the light curve in post-processing with arbitrary time bins without losing the original data, give Aqueye and Iqueye unprecedented capabilities for performing timing studies in the optical band. They have an impressive dynamic range of 6 order of magnitudes, the capability to adjust the post-processing analysis to optimize the S/N ratio for the specific source considered, and the sensitivity to detect tiny relativistic corrections in the photon arrival times.

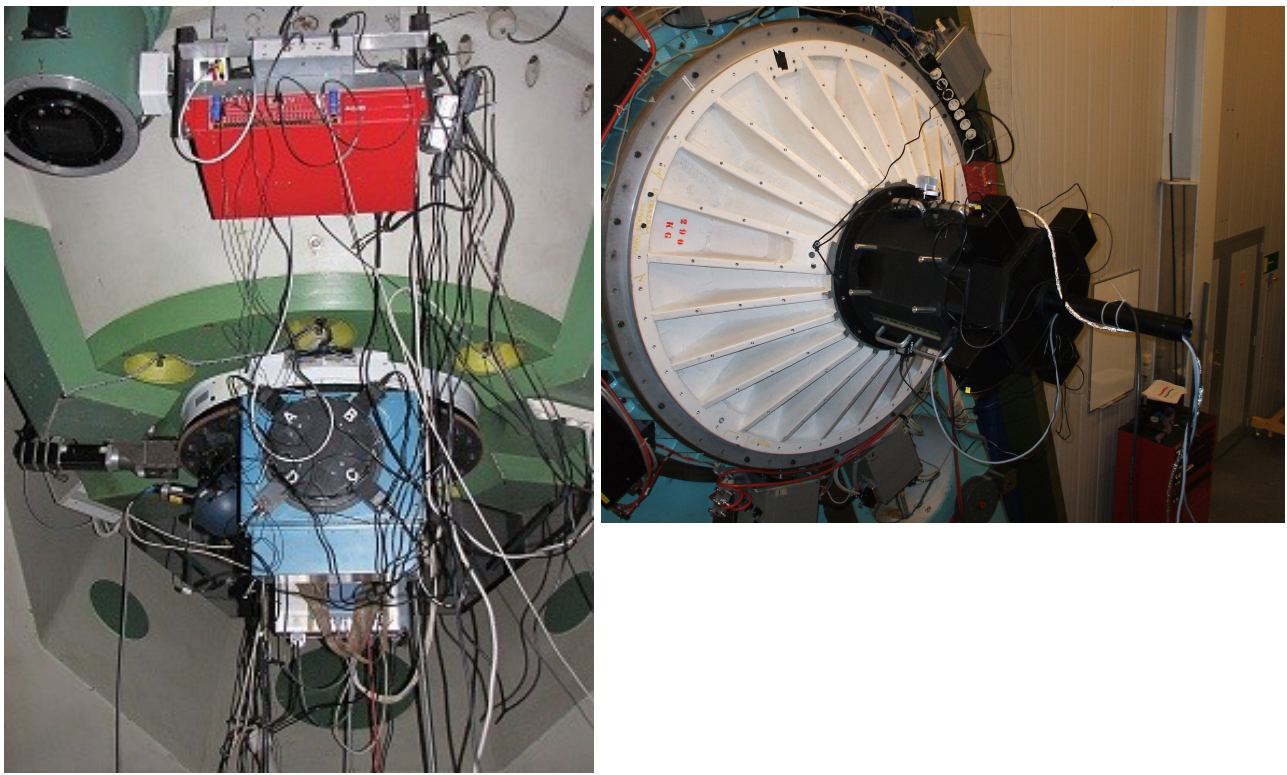

Figure 1. *Left*: Aqueye mounted at the 1.8 m Copernico Telescope at Cima Ekar, Asiago, Italy. Aqueye is the black cylinder-shaped box visible in the image, where 4 SPADs (denoted by A, B, C, D) are attached. *Right*: Iqueye mounted at the 3.6 m ESO NTT telescope in La silla, Chile.

**Optical design**

The entrance aperture of the instrument captures a field of view of ~10 arcsec. Inside the instrument the light beam crosses a focal reducer, where optical elements can be inserted through suitable filter wheels, and is then split in 4 symmetrical arms by a pyramidal mirror[6,7,11] (Figure 2). In each arm, additional filters and polarizers can be inserted in a collimated portion of the beam. The beam is then refocussed on four SPADs, that have a diameter of 100 micrometers on both Aqueye and on Iqueye. Pinholes with different diameters (from 3.5" to 8.7") can be inserted in front of the pyramidal mirror to match the seeing conditions and maximize the signal to noise ratio. The original version of Aqueye did not have the additional SPAD monitoring the sky and the field camera.

The advantages of this split-pupil design are manifold. It allows to: partly recover dead time effects in each SPAD, increase the sustainable count rate, analyze each channel independently and cross-correlate among the different sub-pupils. Thus, counts from the 4 SPADs can be analyzed independently or summed together during the post-processing analysis, depending on the scientific goals. This makes these instruments ideally suited to perform multicolour polarization resolved photon counting. Instrumental polarization coming from the four faces reflections will be characterized through laboratory measurements. We already tested that, when summing counts from all 4 channels, polarization from the pyramid is effectively averaged out[12].

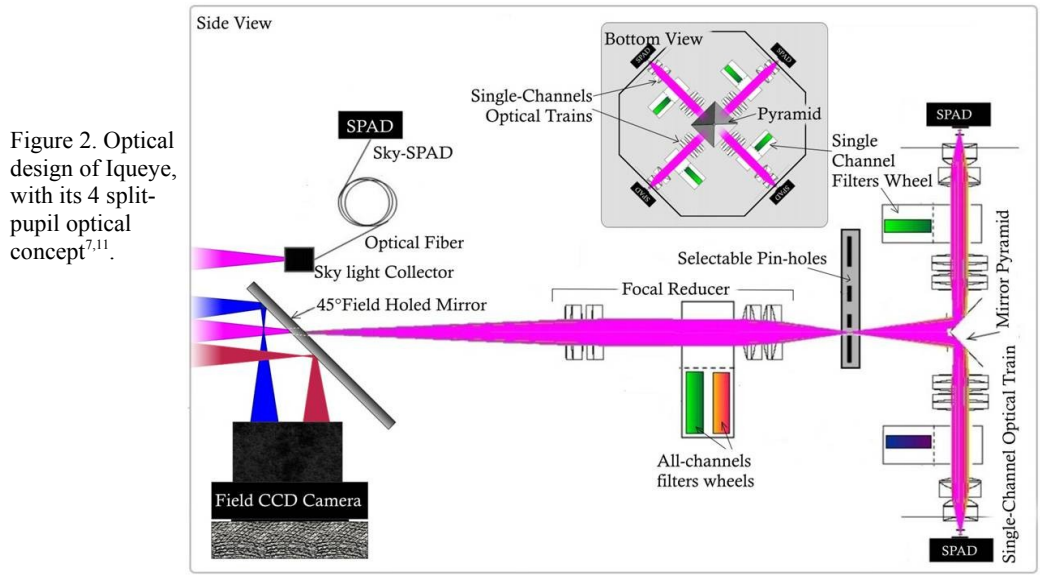

Figure 2. Optical design of Iqueye, with its 4 split-pupil optical concept[7,11].

**Acquisition and timing system**

The signal from the SPADs is sent to a Time To Digital Converter (TDC) board, made by CAEN (Costruzioni Apparecchiature Elettroniche Nucleari, Italy), and then to a dedicated acquisition server[6,7,11] (Figure 3). The TDC makes use of an external Rubidium clock and a GPS unit for checking the long time stability of the clock. The TDC tags each event with a resolution of 24.4 ps per channel and transfers all the data to an external computer through an optical fiber, where the data are acquired. After the end of the observation, data are stored on an external server. SPAD E (that monitors the sky) and the field camera were not present in Aqueye, but are now implemented in Aqueye+.

The system has the capability to time tag and store the arrival time of each individual photon with better than 100 ps relative time resolution (< 0.5 ns absolute time accuracy with respect to UTC[7]).

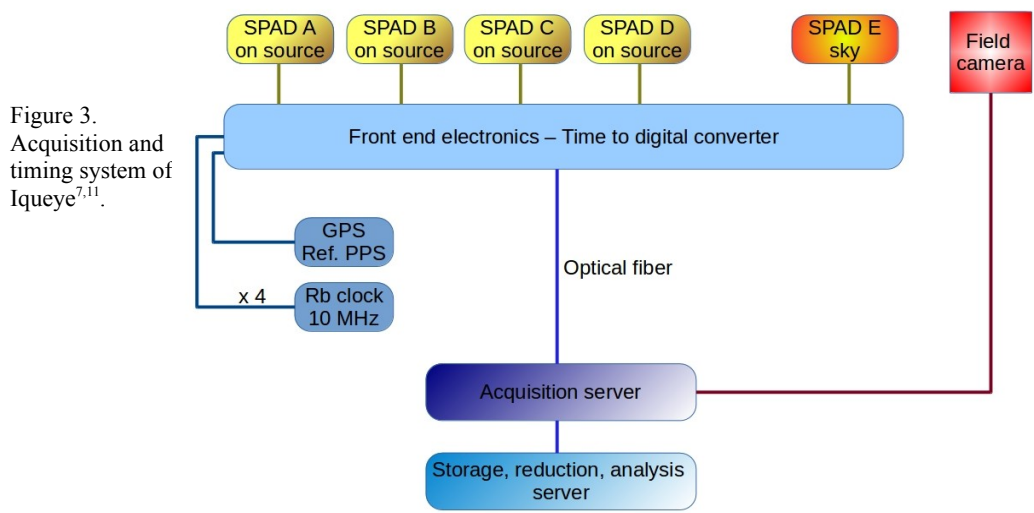

Figure 3. Acquisition and timing system of Iqueye[7,11].

Aqueye and Iqueye have been (and will continue to be) a strong stimulus for the development of several technological areas, in particular for ultra-fast photon counting detectors and high time resolution electronics. Concerning the first aspect, it is worth mentioning our close collaboration with the Italian company MPD, one of the world's leading SPAD manufacturers. The continuous interaction with this company has resulted in significant technological developments, in particular the launch on the market of the first array of SPAD (SPID: Single Photon Imaging Detectors). Albeit these first SPIDs are not yet suitable for HTRA, they are already used in other research areas, as for example in medicine, biology, pharmacology, high-energy physics. Concerning the second aspect, the request of instrumentation capable of performing timing measurements at very high temporal resolution is rapidly increasing: from classical and quantum communication, to laser ranging, to accurate distribution of reference time signals on Earth, to medical equipment for radiotherapy.

## 3. OPTICAL HTRA WITH AQUEYE AND IQUEYE

Optical HTRA concerns itself with observations on time scales of a few seconds or less not normally achievable through conventional cameras using CCDs. Thanks to the aforementioned recent technological developments in photon counting technology, it is an astronomical research field in rapid expansion. We have been among the first to explore applications of modern technology to this field, obtaining results of extraordinary accuracy.

**Optical pulsars**

In the last few years we have successfully used Aqueye and Iqueye to carry out a dedicated observing programme for the timing analysis of optical pulsars. The few known optical pulsars are definitely the most observed and studied objects in the field of HTRA. Although the general pulsar emission model is widely accepted, there are still many open questions that can be solved only through high time resolution observations.

With Iqueye we obtained the best optical light curves ever of the 50 ms pulsar PSR B0540-69 [13] and the 80 ms Vela pulsar[14]. The Crab pulsar, easily observable also from Asiago, has been targeted several times with both Aqueye and Iqueye, obtaining timing information of superior quality. Our typical statistical uncertainty on the determination of the rotational period and of the phase of the main pulse, achieved in an observing run lasting only a few days, are of the order of 0.1 ps and 1 microsecond, respectively[12,15]. Given the capability of assigning a UTC time-tag to each detected photon with an accuracy better than 0.5 ns, a comparative analysis can be made with data acquired by ground radio telescopes, e.g. Jodrell Bank (JB), to investigate the nature of the radio-optical delay[16,17] and the increased visible emission in correspondence of the giant radio pulses[18,19], or those obtained by space telescopes such as Fermi[20]. Comparison with the JB ephemerides shows that the optical pulse in the light curve of the Crab pulsar leads the radio one by 150-250 microseconds[12,15], as shown in Figure 4. This time/phase shift can be caused by a different position of the optical and radio emitting regions (of the order of 50-80 km) but, most probably, the optical and radio beams are misaligned (by 1.5-3°)[12].

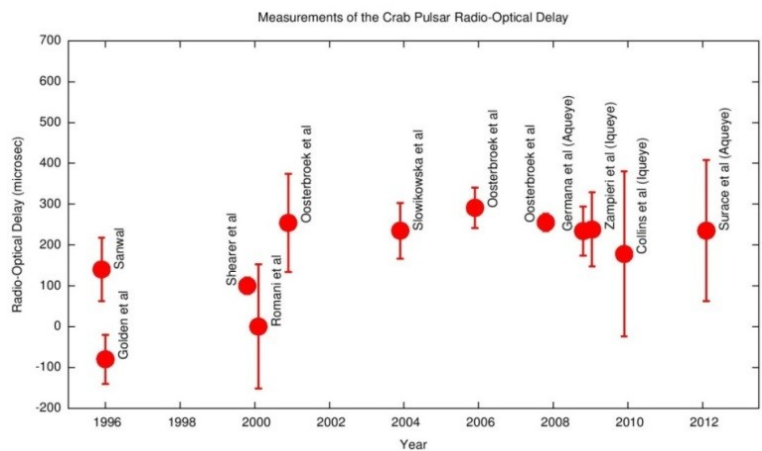

Figure 4. The Crab pulsar radio-optical delay. The error bar in our measurements (2009-2012) is dominated by the radio error, not by the optical one).

With data of such extremely high quality it is also possible to perform an independent phase coherent optical timing analysis of the evolution of the Crab main peak using only Aqueye and Iqueye data. The dispersion around the fit of the timing solution calculated with 2009 Iqueye data is only ~18 microseconds (Figure 5). The strength of the optical data is that they are not at all affected by dispersion measure variations and, in this respect, they offer robust results that provide an independent confirmation of the validity of simultaneous radio timing solutions.

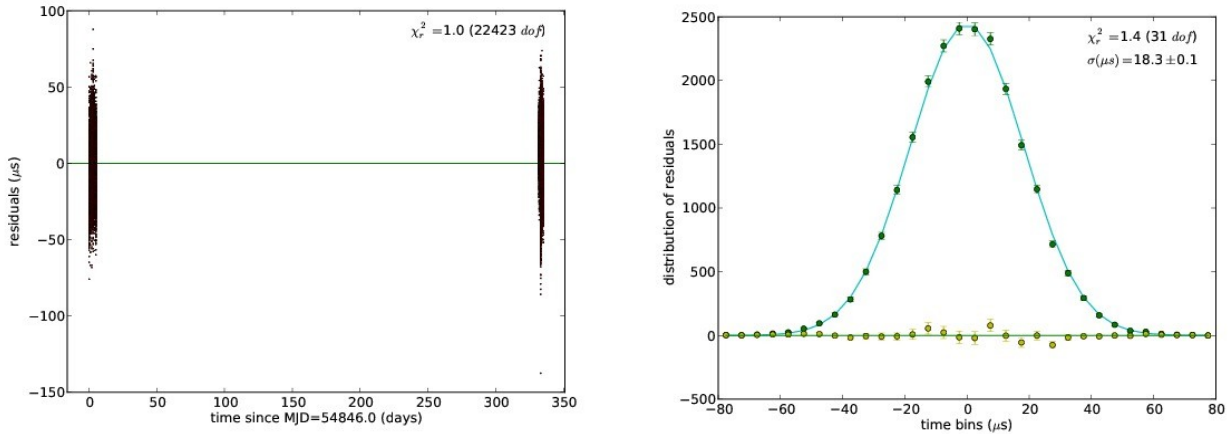

Figure 5. Residuals of the optical timing solution (cubic polynomial in phase) for all the 2009 Iqueye observations of the Crab pulsar (*left*) and their distribution (*right*). A gaussian fit of the distribution with a dispersion of 18 microseconds (solid line) is also shown, along with its residuals at the bottom.

**Transits and occultations**

Another very interesting field of investigation for HTRA are rapid transits of celestial objects, such as near Earth objects, and/or phenomena in which flux variability is not directly generated by the light source but by an intervening body, such as exoplanet transits and occultations. The time resolution requirements for the detection of a single transit are not stringent. However, high time resolution observations become essential when one wishes to identify a third orbiting body, or differences in fast fluctuations during the initial and final part of the transit induced by the planetary atmosphere. Similar considerations apply to planet/asteroids/Kuiper belt objects stellar occultations. Consider as an example the stellar occultation by the trans-neptunian object Makemake[21]. The disappearances and reappearances of the occulted star were abrupt, showing that Makemake has no global Pluto-like atmosphere. However, a short spike (~few seconds) seen in the light curve in the middle of the occultation may, in fact, suggest the presence of refracted light by a local atmosphere somewhere on the limb of the planet.

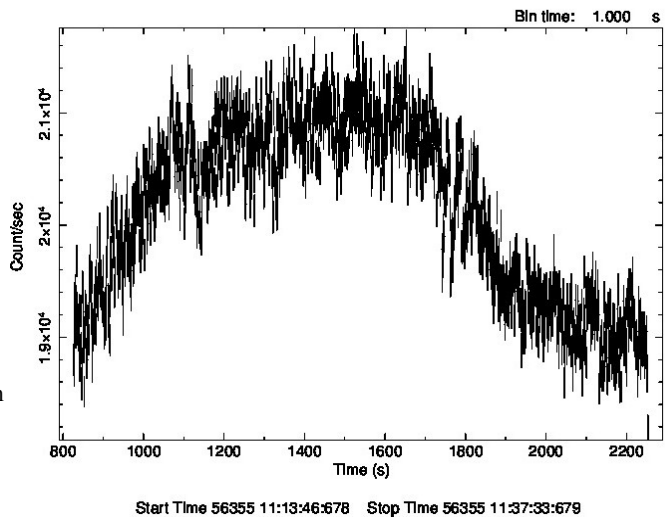

Figure 6. Light curve of the Aqueye observation of the passage of asteroid 2283 Bunke close to star USNO-B1.0 0858-00222444 (March 3, 2013 – from 23:13:46.7 to 23:37:33.7 UT).

A related phenomenon that we directly attempted to observe was the quasi-occultation of the asteroid 2283 Bunke that, on the 3$^{rd}$ of March 2013, was expected to pass very close to and possible occult the field star USNO-B1.0 0858-00222444. The light curve of the passage taken with Aqueye is shown in Figure 6. During passage, the instrument was steadily centered on the field star. As the asteroid was entering the SPADs field of view, the intensity raised. Then, it remained constant for approximately 500 seconds, during which both the field star and the asteroid were inside the field of view. Finally, it decayed as the asteroid moved out of view. Although the asteroid did not occult the star during this passage (no sharp decrease and increase of the flux was observed during maximum light), this observation showed all the potential of Aqueye for high time resolution observations of events of this type.

## 4. NEW TECHNOLOGICAL FEATURES IMPLEMENTED ON AQUEYE+

Thanks to two major grants from the University of Padova and the fondazione CARIPARO in Padova, during 2013-2014 we undertook a major upgrade of Aqueye that ended up, in fact, in the construction of a largely new instrument, Aqueye+. *The primary object of this project was to build new opto-mechanics and an optimized optical train, to improve the photometric performances of the instrument and to make it remotely controlled, easily mountable and operational at the Copernicus telescope in Asiago.* Achieving this goal has made Aqueye+ not only the most powerful astronomical photometer in terms of dynamic range and time resolution, with multi-colour and/or multi-polarization capabilities, but also a more flexible instrument requiring significantly reduced technical operations.

**New opto-mechanical design and an optimized optical train**

One of the main requirements deriving from the global reconfiguration of the Asiago Observatory, managed by INAF-Astronomical Observatory of Padova, was to make Aqueye a fully remotely controlled instrument, completely independent of external opto-mechanical devices. The remote control is meant to minimize the human and operational resources needed to manage the observing site. Aqueye was relaying entirely on the optical and mechanical train of the Asiago Faint Object Spectrograph and Camera (AFOSC[22]), a low resolution spectrograph and imager mounted at the Copernico telescope (the blue box in Figure 1, left).

The new instrument has a new optical design and and independent optical bench, with its own opto-mechanics and telescope flange (see Figures 7 and 8).

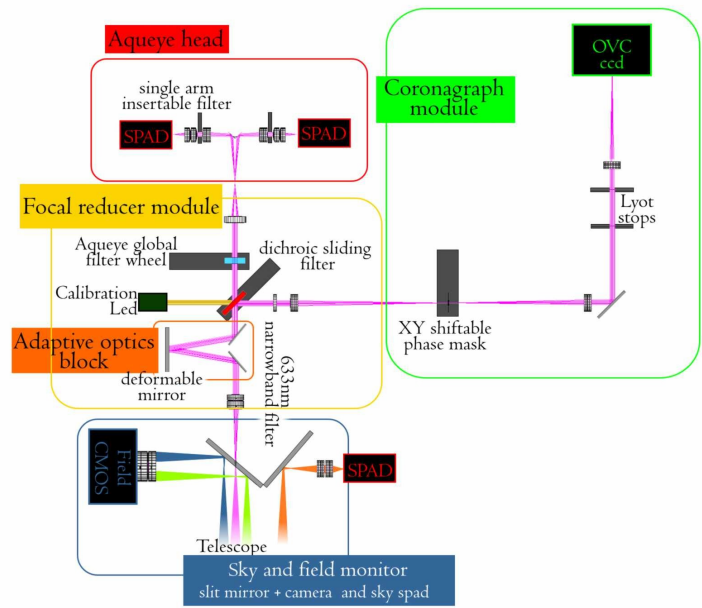

Figure 7. Optical design of Aqueye+. The coronographic module is under construction and is not currently part of the instrument. Similarly, the adaptive optics module is not currently implemented.

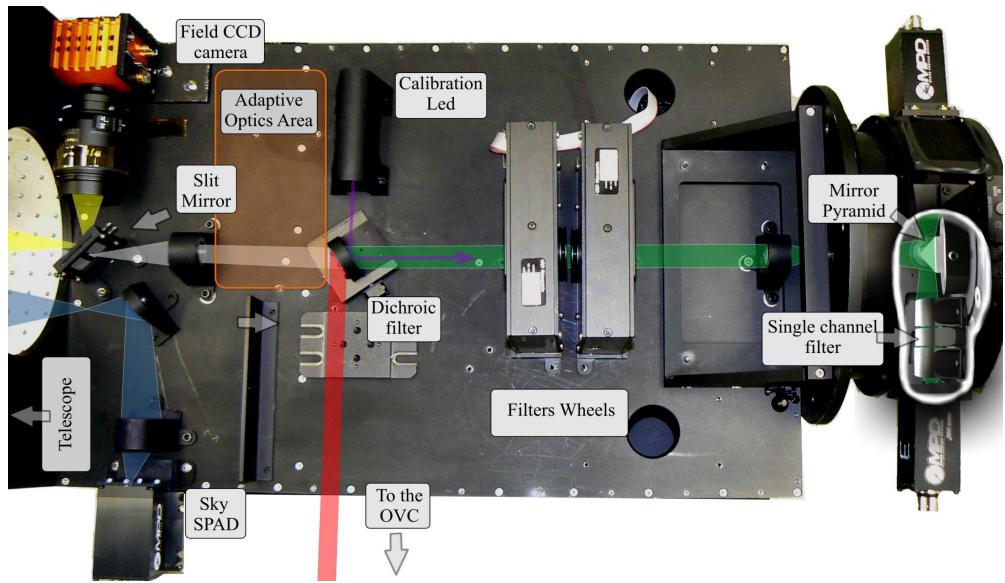

Figure 8. Optical bench of Aqueye+. On the right the optical module with the piramid and MPD SPADs, belonging to Aqueye. On the left the new optical system, with the optical elements and mirrors, and the Osprey Raptor Photonics field camera (the bright orange box on the image). The coronographic module is not present.

The optical design is very similar to that of Iqueye, with the addition of a dichroic sliding filter and a calibration LED controlled by an external circuit[11,23] (Figure 7). The coronographic module and adaptive optics module are under construction[24] and not currently implemented. They are not presented in this paper. The adaptive optics module and dichroic sliding filter are needed mainly for the additional coronographic module, although the former will also increase the overall optical efficiency of the system and the latter is used also to insert the beam from the calibration LED in the optical path. The optical train of the focal reducers is completely new and optimized for this optical design. The mirror in front of the optical beam has a narrow slit. Most of the telescope field of view is redirected towards the field camera. An additional SPAD is monitoring the sky thanks to a smaller mirror placed nearby, that is intereecepting part of the entrance field of view.

**High sensitivity and high frame rate field camera**

One of our primary goals in developing Aqueye+ was to improve its photometric calibration. The few arcsec field of view of Aqueye has shown some limitations when observing non-periodic variable objects. In fact, it was not possible to monitor reference stars and perform differential photometry. To realize a very accurate differential photometry on short timescales, we installed on Aqueye+ a CMOS high sensitivity and high frame rate camera, that monitors reference stars in the field of view to measure intensity fluctuations due to atmospheric seeing and sky conditions[11,23]. Our final choice fell on a Osprey sCMOS Raptor Photonics camera (the bright orange box on the image in Figure 8), featuring a 2kx2k low noise scientific CMOS sensor that stands out with its extreme sensitivity, high resolution, high speed and excellent Quantum Efficiency. The camera is peltier cooled, uses a 12 bit A/D converter and offers a standard CameraLink output.

## 5. NEW/UPGRADED SOFTWARE TOOLS IMPLEMENTED ON AQUEYE+

**Re-designing the acquistion and control systems and operating remotely**

The new acquisition and timing system of Aqueye+ resembles that of Iqueye (Figure 3). However, to make Aqueye+ a fully integrated instrument for HTRA, it is crucial to jointly manage the acquisition from the SPADs and the new field camera. To this end, we bought a new powerful workstation (aqkite; Dell Inc. Precision T5600 Intel with a 8 cores Xeon processor E5-2600 and a NVIDIA Tesla C2075 GPGPU) and re-designed the whole acquisition software, porting it to the new 64-bit architecture. In addition, to operate remotely Aqueye+ from the more easily accessible Pennar station required some changes in the instrument hardware and the realization of a suitable software interface to control the instrument sub-systems (GPS, filter wheels, dichroic filter slit, temperature/humidity sensor). The first successful attempts to observe remotely have been done in the last observing run of January 2015.

A screenshot of the user interface to the new Aqueye+ acquisition system (osserva[25]) is shown in Figure 9. It allows the user to launch and manage simultaneously the acquisition of events from the CAEN TDC through a dedicated 64-bit acquisition software developed by our team (acquire v. 20140731) and the images from the field camera through the commercial software package EPIX XCAP v. 3.8 (http://www.epixinc.com/products/xcap.htm). All parameters of the acquisition (including the length of the observation, the buffer size of the acquired data segments, the duration of exposures, the camera capture rate, the filters inserted) are saved in a log file.

Figure 9. Screenshot of the user interface to the new Aqueye+ acquisition system.

**Reduction and analysis software packages**

Raw data and images are saved and, later on, transferred to a storage server (controllo) with 20 TeraBytes of dedicated disc memory space. On this server we perform the post-processing reduction and analysis of the data.

The reduction of the CAEN TDC signal is done with a proprietary software (quest v. 1.1.5[26]), that time tags the events in the raw data files referring them to UTC. The present version of this software contains a few fixes that improve the accuracy of the time tags and a new algorithm that checks the correct alignment to UTC of the first pulse-per-second after the start of the observation. In the near future the GPS unit disciplining the rubidium clock will also be replaced to further improve the time tagging accuracy of the system. If needed, time tags are referred to the barycenter of the Solar System through the standard software package TEMPO2[27,28]. Data are then processed through a dedicated python pipeline[29], that performs different automatic tasks and produces a number of standard output files for the following timing analysis (light curves, power spectra, detection significance, refined search for frequency peaks in the power spectrum, folded profile). Sample outputs of the SPAD signal at the end of the reduction-analysis chain are shown below.

The reduction of the field camera images is done with a python interface to astropy (http://www.astropy.org/), a dedicated module for astronomical analysis that allows to perform standard manipulation and analysis of astronomical images in FITS format. Specific tasks for non-standard analysis of the images, as for example an automatic procedure for real time relative astrometry between a field star and a moving object (like an asteroid) in a sequence of images, have been developed. We are also implementing an automatic cross-calibration procedure, using sequence of images taken during acquisition with the SPADs, to calibrate photometrically the SPAD rates[30].

## 6. INSTRUMENT COMMISSIONING AND CALIBRATION

The first transportation and assembly of Aqueye+ at the Copernico telescope was done on the 2$^{nd}$ and 3$^{rd}$ of May 2014 (see Figure 10). Two technical runs (May and July 2014) were completely dedicated to solving various technical issues and to optimizing the instrument performances. The first light and scientific data taking was done in the following observing run (Nov 2014). On the 19$^{th}$ and 20$^{th}$ of November, we acquired the first Aqueye+ observations of the Crab pulsar in white light and with filters on arms (Figure 11), and the first real time sequence of images of a star (69 Geminorum) with the field camera (Figure 12).

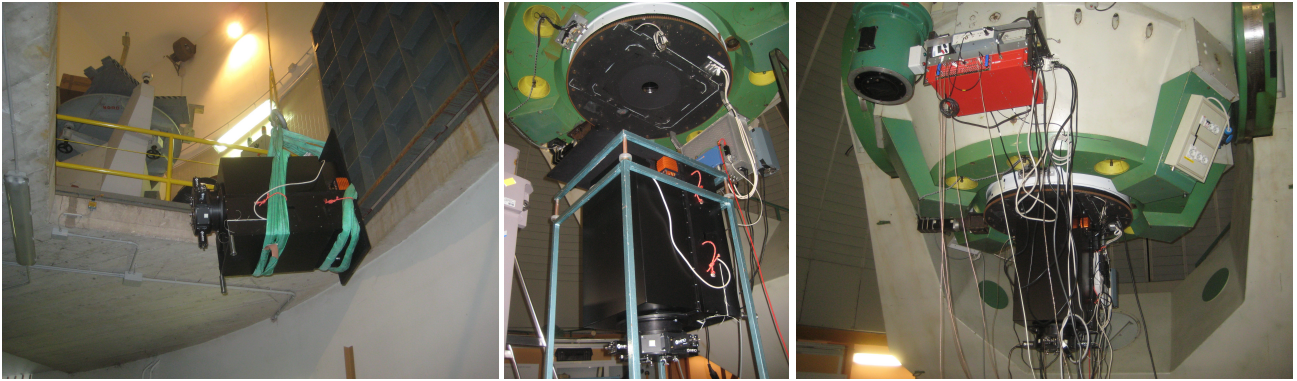

Figure 10. First assembly of Aqueye+ at the Copernico telescope, Cima Ekar, Asiago, on th 3$^{rd}$ of May 2014. *Left*: Lifting to the dome floor. *Center*: Mounting at the telescope. *Right*: Aqueye+ after completing the assembly of the instrument and electronics.

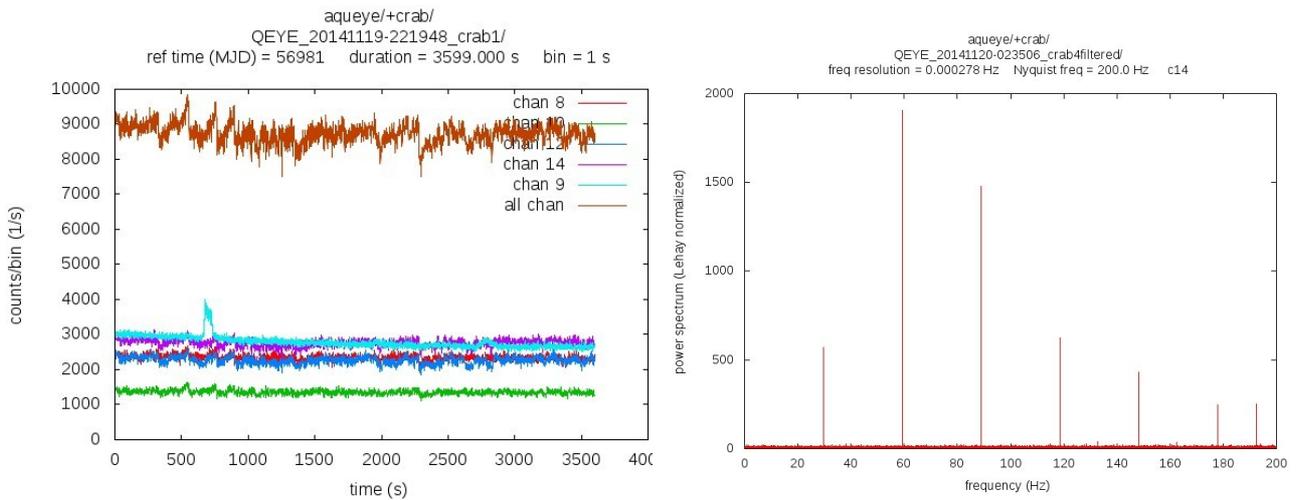

Figure 11. *Left*: Light curve of the first Aqueye+ observation of the Crab pulsar in white light taken on Nov 19, 2014. The bin time is 1s. Count rates on each SPAD and on the sum on all on-source channels are shown. Differences among the 4 SPADs are caused mostly by a non-perfect centering of the source inside the pinhole. Channel (SPAD) 9 is monitoring the nearby sky and shows the passage of an object inside the field of view after ~600-700 s from the start of the observation. *Right*: Unaveraged power spectrum of an Aqueye+ observation of the Crab pulsar taken with a blue filter on Nov 20, 2014. The bin time is 2.5 ms. The peak at ~190 Hz is an aliasing of the seventh harmonic. (*Sample outputs from the Aqueye+ timing analysis pipeline*).

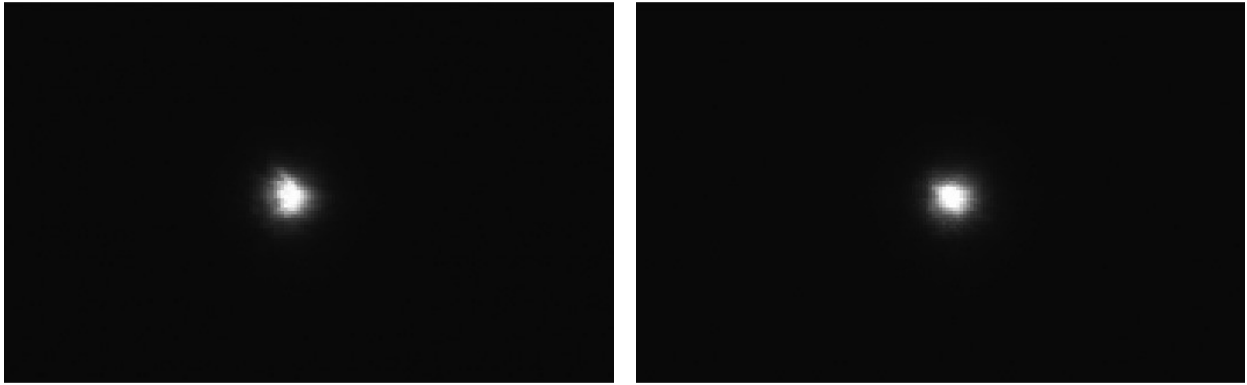

Figure 12: First two images (from a 60 s sequence) taken with the Aqueye+ field camera on Nov 19, 2014 (start time 05:48:50.6). The exposure time is 10 ms and the capture time interval 0.1 s.

To perform an independent test of the linearity and frequency/timing calibration of our acquisition system, and its long term stability, we designed and realized a digital calibration circuit that controls a coloured (RGB) LED lamp inside the instrument[11] (Figure 7). The light beam from the LED is inserted into the optical path through a dichroic sliding filter. The clock of the circuit can be controlled by both an internal H14 Oscillator and by the external Rubidium clock. We performed extensive tests during both runs of Nov 2014 and Jan 2015. Figure 13 shows an enlargement around the maximum frequency detected in the power spectrum of an approximately square wave input signal at ~1.2 kHz. Both signals are generated by 12 binary divisions of the clock input reference frequency which is 4.915 MHz for the internal H14 Oscillator and 5 MHz (to less than 0.005 part per million) for the external Rubidium clock[11]. Therefore, the expected frequencies of the signal are $4915000/2^{12}=1200$ Hz in the first case and $5000000/2^{12}=1220.703$ Hz in the second case, in perfect agreement with the detected frequencies (considering the frequency resolution of 0.017 Hz). A similar test performed in the last run of Jan 2015 shows also that the frequency has remained stable. Similar results are obtained also with input signals of smaller frequency and with the other colours.

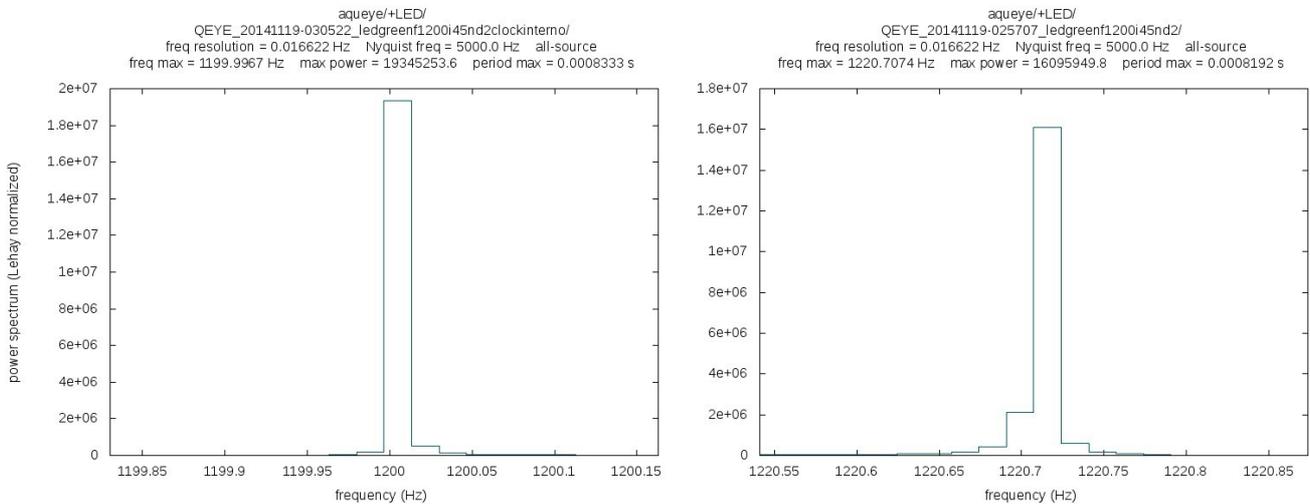

Figure 13: Enlargement around the maximum frequency of the power spetrum of an approximately square wave input signal at ~1200 Hz. *Left*: signal generated with the internal H14 Oscillator. *Right*: signal generated with the external Rubidium clock.

We tested also the linear response of the SPADs and the whole acquisition system with signals in the range from 0.6 Hz to 2.4 kHz.

Among the future planned calibration activities, there is the characterization of the polarimetric response of Aqeye+ using available software tools. This will allow us to use Aqeye+ for calibrated simultaneous polarimetric observations of the Crab pulsar. We will check it on optical bench with a polarized calibration source. In addition, we will perform an accurate instrumental characterization of the field camera.

## 7. SCIENTIFIC HIGHLIGTHS

We performed the first two scientific runs with Aqeye+ mounted at the Copernico telescope in November 2014 and January 2015. A precise observing procedure has to be followed for correctly operating the telescope and instrument remotely from the Pennar station[31].

**Simultaneous multi-colour observations of the Crab pulsar**

In the observing runs of Nov 2014 and Jan 2015 we successfully performed simultaneous multi-colour observations of the Crab pulsar in three different optical bands (R, V, B). In the Nov 2014 run, the quality of the data was very good thanks also to the optimal seeing conditions (< 1.2"). In Figure 14 we show the folded profile acquired simultanesouly in the three bands of one observation taken on November 20. A very preliminar analysis shows that the peak of the folded profile in the three bands are aligned within ~100 microseconds. A more detailed analysis, including all the available data of the observing run, is currently underway.

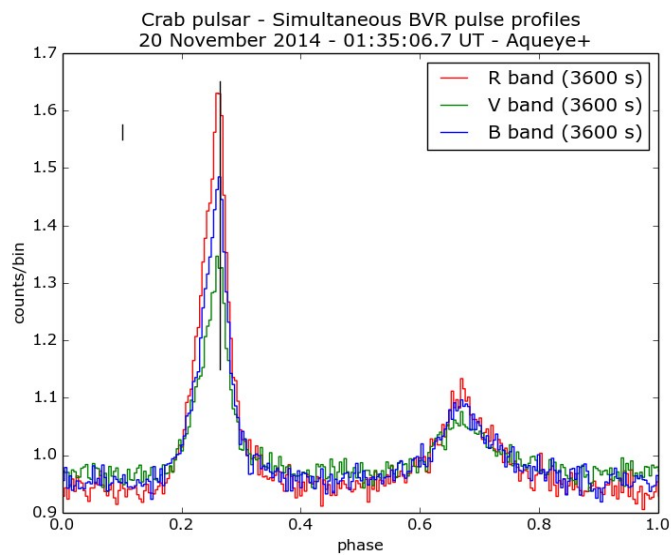

Figure 14: Simultaneous pulse profiles in the BVR optical bands obtained with Aqeye+ in November 2014. The phase bin is 1/256, corresponding to 130 microseconds. The long vertical line marks the phase of the maximum in the B band, while the short line in the top-left corner denotes the typical error bar.

We plan to continue to perform these measurements and also to make simultaneous observations with X- and gamma-ray observatories. The large availability of telescope time is essential for observing transient non-periodic events, as attested by the occurrence of the 2013 X- and gamma-ray flare of the Crab pulsar when Aqeye in Asiago was the only instrument providing simultaneous optical data[32].

**Relative timing astrometry of the asteroid 2004 BL86**

On 27 January 2015 we observed the transit of the asteroid 2004 BL86 during its near-Earth passage. We took four sequence of images with the Aqeye+ field camera starting at 02:15:06 UT and ending at 02:39:35 UT. The frame rate is 7.7 frames/s and each frame has an exposure time of 0.1 s. Each sequence lasts approximately 90 seconds (600-700 frames). The asteroid passed close to 3 field stars. In particular, it almost occulted star TYC 809-1498-1

(RA=08:35:27.432 DEC=+13:50:27.54 , J2000). A movie of this transit is available at https://www.youtube.com/watch?v=M-mGOjV1AV8&list=UL (Newsletter Media INAF of Feb 3, 2015). Figure 15 (left) shows the image corresponding to the time of closest approach. Accurate timing astrometry of the images yields the following time and angular distance at closest approch: 02:30:05.8 UT, 3.2" (Figure 15, right). Errors are dominated by systematics related to the combined effects of seeing and telescope guiding and are conservatively estimated to be 1.5 s in time and 3" in position. A more accurate analysis is underway. The plate scale of the field camera is 0.11 "/pixel. This type of information is important to constrain the orbital parameters of these potentially hazardous objects.

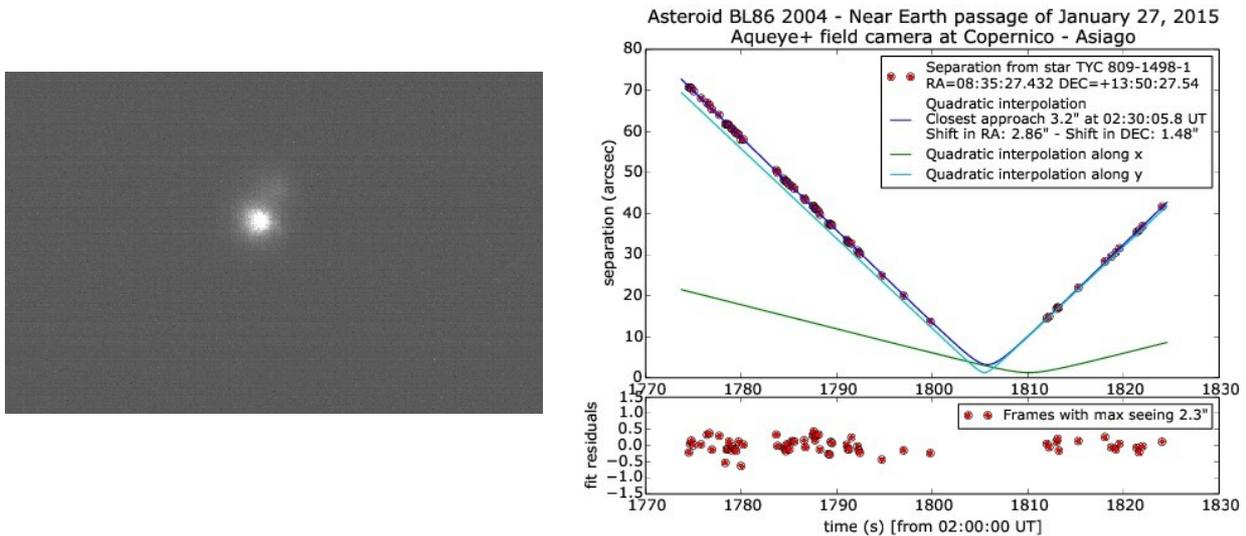

Figure 15: Closest approach of asteroid 2004 BL86 to star TYC 809-1498-1 (RA=08:35:27.432 DEC=+13:50:27.54 , J2000, magnitude V=11.4) during its near-Earth passage on 27 January 2015, as seen with the Aqueye+ field camera. *Left*: snapshot of the image sequence at closest approach. The asteroid is the brigtest object in the image. *Right*: relative timing astrometry of the image sequence, with the results of the best fitting parabola. The time and angular distance at closest approch are: 02:30:05.8 UT, 3.2".

## 8. CONCLUSIONS AND FUTURE DEVELOPMENTS

During in 2013-2014 we built Aqueye+, a new ultrafast optical single photon counter, based on SPAD detector technology and a 4-fold split-pupil concept. It is a major upgrade of its predecessor Aqueye. *The primary object of this project was to build new opto-mechanics and an optimized optical train, to improve the photometric performances of the instrument and to make it remotely controlled, easily mountable and operational at the Copernicus telescope in Asiago.*

We have presented the new technological features implemented on Aqueye+, namely a state of the art timing system, a dedicated and optimized optical train, a high sensitivity and high frame rate field camera and remote control, which will give Aqueye plus much superior performances with respect to its predecessor, unparalleled by any other existing fast photometer. In the near future the instrument will host also an optical vorticity module to achieve high performance astronomical coronography and a real time acquisition of atmospheric seeing unit[23]. Future plans are to permanently mount it at the Nasmyth focus, which will insure maximum operational flexibility and minimal operational intervention. With these characteristics Aqueye+ is not only the most powerful astronomical photometer in terms of dynamic range and time resolution, with multi-colour and/or multi-polarization capabilities, but also a flexible instrument requiring significantly reduced technical operations.

We presented also some highlights of the scientific results obtained with Aqueye+ in the last two scientific observing runs performed in November 2014 and January 2015: the first simultaneous multi-colour (BVR) observations of the Crab pulsar and the relative timing astrometry of the transit of asteroid 2004 BL86 during its near-Earth passage on 27 January 2015. The observational and scientific activity is only at the beginning. We will perform multicolour polarization-resolved photometry of the Crab pulsar and some of the most rapidly variable objects in the Universe, that we started to study with Aqueye and Iqueye, such as pulsating white dwarfs, cataclysmic variables, stellar flares. Short timescale variability in these objects is also an area of primary interest in optical HTRA. For this, we will take full advantage of the photometric calibration performed with the high sensitivity field camera. When the instrument will be mounted at the Nasmyth focus, it will allow us to frequently monitor the Crab pulsar and the difference of the pulse arrival time in different spectral bands. It will be possible also to implement a more flexible Target of Opportunity (ToO) observing mode, for targeting interesting transients and the Crab pulsar at the time of gamma-ray flares.

The successful observations of the asteroid 2004 BL86 shows the feasibility and accuracy with which transits and occultation events can be followed (fast events, like 2004 BL86, with the field camera while slower events, like 2283 Bunke, with the SPADs). During the occultation event produced by Makemake[21], the star was around the 18$^{th}$ magnitude, well within the sensitivity of Aqueye+ even in the multicolour polarization resolved photometric mode. In addition, with a time resolution better than milliseconds, it is possible to go over a spatial accuracy of milli-arcsec.

Aqueye+ is a further step in the development of a new generation of photometers that will lead to a significant advancement in our understanding and ability to investigate the Universe. This is of fundamental importance also in view of the construction of the new extremely large telescopes (e.g. the ESO E-ELT), for which the interest of the international HTRA community is very high. Such developments will be of great interest also in connection with applications connected to Intensity Interferometry[33], especially for the forthcoming Cherenkov Telescope Array and INAF-ASTRI Mini-array.

## 9. ACKNOWLEDGMENTS

We would like to thank all the staff of the Asiago Cima Ekar observing station for their continuous support and effective collaboration, essential for the success of this project. We also thank Mirco Zaccariotto and his collaborators for helping in the realization of the opto-mechanical design, Tommaso Occhipinti for developing the adaptive optics module, and Marco Fiaschi for adapting the software interface for telescope pointing and guiding to the new instrumental setup and for developing part of the sub-system control software. This work is based in part on observations collected at the Copernico telescope (Asiago, Italy) of the INAF-Osservatorio Astronomico di Padova under program "Timing ottico della Crab pulsar ad elevatissima risoluzione temporale con Aqueye+". This research has been partly supported by the University of Padova under the Quantum Future Strategic Project, by the Italian Ministry of University MIUR through the programme PRIN 2006, by the Project of Excellence 2006 Fondazione CARIPARO, and by INAF-Astronomical Observatory of Padova under the grant "Osservazioni con strumentazione astronomica ad elevata risoluzione temporale e modellizzazione di emissione ottica variabile".

## REFERENCES

[1] Dhillon, V. S. et al., "ULTRACAM: an ultrafast, triple-beam CCD camera for high-speed astrophysics", MNRAS, 378, 825 (2007).

[2] Collins, P. et al., "GASP - Galway Astronomical Stokes Polarimeter", in Proceedings of the Conference 'Polarimetry days in Rome: Crab status, theory and prospects', held in Rome, 16-17 October 2008, Proceedings of Science (2009). arXiv:0905.0084

[3] Kanbach, G. et al., "OPTIMA: A High Time Resolution Optical Photo-Polarimeter", ASSL, 351, 153 (2008).

[4] McPhate, J. B. et al., "BVIT: A Visible Imaging, Photon Counting Instrument on the Southern African Large Telescope for High Time Resolution Astronomy", Physics Procedia, 37, 1453 (2012).

[5] Mazin, B. A. et al., "ARCONS: A 2024 Pixel Optical through Near-IR Cryogenic Imaging Spectrophotometer", PASP, 125, 1348 (2013).

[6] Barbieri, C. et al., "AquEYE, a single photon counting photometer for astronomy", J. Mod. Opt., 56, 261 (2009).

[7] Naletto, G. et al., "Iqueye, a single photon-counting photometer applied to the ESO new technology telescope", A&A, 508, 531 (2009).

[8] Cova, S. et al., "Evolution and prospects for single-photon avalanche diodes and quenching circuits", J. Mod. Opt., 51, 1267 (2004).

[9] Billotta, S. et al., "Characterization of detectors for the Italian Astronomical Quantum Photometer Project", J. Mod. Opt., 56, 273 (2009).
[10] Dravins, D, et al., "QuantEYE: The Quantum Optics Instrument for OWL", Proceedings from meeting 'Instrumentation for Extremely Large Telescopes', held at Ringberg Castle, July 2005, T.Herbst, ed. (2005). astro-ph/0511027
[11] Zampieri, L. et al., "Aqueye/Iqueye technical documentation", Internal Technical Report (2015).
[12] Zampieri, L. et al., "Optical phase coherent timing of the Crab nebula pulsar with Iqueye at the ESO New Technology Telescope", MNRAS, 439, 2813 (2014).
[13] Gradari, S. et al., "The optical light curve of the Large Magellanic Cloud pulsar B0540-69 in 2009", MNRAS, 412, 2689 (2011).
[14] Verroi, E. et al., "Vela pulsars observations with Iqueye mounted at the ESO New Technology Telescope", in preparation.
[15] Germanà, C. et al., "Aqueye optical observations of the Crab Nebula pulsar", A&A, 548, A47 (2012).
[16] Oosterbroek, T., et al., "Absolute timing of the Crab Pulsar at optical wavelengths with superconducting tunneling junctions", A&A, 456, 283 (2006).
[17] Oosterbroek, T. et al., "Simultaneous absolute timing of the Crab pulsar at radio and optical wavelengths", A&A, 488, 271 (2008).
[18] Shearer, A. et al., "Enhanced Optical Emission During Crab Giant Radio Pulses", Science, 301, 493 (2003).
[19] Strader, M. J., et al., "Excess Optical Enhancement Observed with ARCONS for Early Crab Giant Pulses", ApJ, 779, L12 (2013).
[20] Abdo, A. A. et al., "Fermi Large Area Telescope Observations of the Crab Pulsar And Nebula", ApJ, 708, 1254 (2010).
[21] Ortiz, J, L. et al., "Albedo and atmospheric constraints of dwarf planet Makemake from a stellar occultation", Nature, 491, 566 (2012).
[22] Tomasella, L. et al., "Afosc @1.82 m Copernico Telescope, Ekar - User Manual" (2015). www.pd.astro.it/images/pdf_asiago/manuale_afosc.pdf
[23] Naletto, G. et al., "Aqueye Plus: a very fast single photon counter for astronomical photometry to quantum limits equipped with an Optical Vortex coronagraph", Proc. SPIE 8875, id. 88750D (2013)
[24] Verroi, E. et al., "Aqueye+: a wavefront sensorless adaptive optics system for narrow field coronagraphy", Proc. SPIE 8864, id. 88641W (2013).
[25] Barbieri, M., "Aqueye/Iqueye acquisition and control system", Internal Technical Report (2015).
[26] Zoccarato, P., "Quest User Manual", Internal Technical Report (2015).
[27] Hobbs, G.B., et al., "TEMPO2, a new pulsar-timing package - I. An overview", MNRAS, 369, 655 (2006).
[28] Edwards, R. T., et al., "TEMPO2, a new pulsar timing package - II. The timing model and precision estimates", MNRAS, 372, 1549 (2006).
[29] Zampieri, L., "Aqueye/Iqueye timing analysis pipeline", Internal Technical Report (2015).
[30] Zampieri, L., "Aqueye/Iqueye field camera pipeline", Internal Technical Report (2015).
[31] Zampieri, L., "Aqueye+ Observing Guide", Internal Technical Report (2015).
[32] Zampieri, L. et al., "Optical Observations of the Crab pulsar from March 2 through 4, 2013, with Aqueye at the Copernico telescope in Asiago (Cima Ekar Observatory)", ATel #4878 (2013).
[33] Dravins, D., et al., "Optical intensity interferometry with the Cherenkov Telescope Array", Astropart. Phys., 43, 331 (2013).